\documentclass[12pt]{article}
\usepackage{epsf,amsfonts}

\epsfverbosetrue
\def\double{\mathbb}         
\def\ccal{\cal}           

\def\cc{{\double C}}

\def\zz{{\double Z}}

\def\aa{{\cal A}}

\def\dd{{\cal D}} 
        
\def\hh{{\cal H}}
\def\hhh{{{\double H}}}   

\def\mm{{{\ccal M}}}

\def\aa{{\cal A}}
\def\dd{{\cal D}} 
\def\hh{{\cal H}}

\def\lll{{\cal L}}

\def\t{{\rm tr}\,}

\def\ddd{{\,\hbox{$\partial\!\!\!/$}}}

\def\ot{\otimes}
\def\op{\oplus}

\def\bb{\begin{eqnarray}}
\def\ee{\end{eqnarray}}
\def\eee{\nonumber\end{eqnarray}}
\def\pp{\pmatrix}
\def\qq{\quad}

\begin{document}

\hsize 17truecm
\vsize 24truecm
\font\twelve=cmbx10 at 13pt
\font\eightrm=cmr8
\baselineskip 18pt

\begin{titlepage}

\centerline{\twelve CENTRE DE PHYSIQUE TH\'EORIQUE}
\centerline{\twelve CNRS - Luminy, Case 907}
\centerline{\twelve 13288 Marseille Cedex 9}
\vskip 3truecm

\centerline{\twelve STANDARD MODEL AND
UNIMODULARITY CONDITION}

\bigskip

\begin{center}
{\bf Serge LAZZARINI,
\footnote{\, also at Universit\'e
de la M\'editerrann\'ee\qq sel@cpt.univ-mrs.fr} 
 Thomas SCH\"UCKER, 
\footnote{\, also at
Universit\'e de Provence \qq 
schucker@cpt.univ-mrs.fr }}\\

\end{center}

\vskip 2truecm
\leftskip=1cm
\rightskip=1cm
\centerline{\bf Abstract} 

\medskip

The unimodularity condition in Connes' formulation
of the standard model is rewritten in terms of group
representations. 
 
 \vskip 1truecm\begin{center}
 Contribution to the Workshop on Quantum Groups\\
 Quantum Groups, Noncommutative Geometry\\
 and
Fundamental Physical Interactions\\
ICI Guccia, Palermo, Italy\\
December 12--17 1997
\end{center}
\vskip 1truecm
PACS-92: 11.15 Gauge field theories\\ 
\indent
MSC-91: 81T13 Yang-Mills and other gauge theories 
 
\vskip 1truecm

\noindent january 1998
\vskip 1truecm
\noindent CPT-98/P.3606\\
\noindent hep-th/yymmxxx
 
\vskip1truecm

 \end{titlepage}

\section{Introduction}
 
Is our universe open or closed? It is likely that we
will never have an experimental answer to this old
question. Due to the small value of the speed of light
the global geometry of spacetime escapes observation.
On the other hand the global properties of the
rotation group are well established in quantum
physics. Neutrons have to be rotated through an
angle of $720^{\rm o}$  before interference patterns
repeat
\cite{neu}. Mathematically this means that spin
${\textstyle\frac{1}{2}}\ $ particles are represented
only up to a phase under the rotation group $SO(3)$.
If we want genuine representations we must use its
universal cover $SU(2)$. In the same spirit we may
wonder about the global nature of the internal group
$SU(2)\times U(1)\times SU(3)$ coding electro-weak
and strong forces. It is well known \cite{or} that the
representations classifying quarks, leptons, gauge and
Higgs bosons in the standard model are already
representations of $S(U(2)\times U(3))$. This
experimental fact has no explanation in the frame of
Yang-Mills-Higgs theories. In the frame of Connes'
noncommutative geometry \cite{connes}, where
internal groups do not fall from heaven but are
derived from associative involution algebras, this
coincidence is vital. 

\section{The global nature of the standard model}

For every integer greater than one, $n\ge 2$, the
maps
\bb U(n) & \longrightarrow & SU(2)\times U(1)\cr 
u\ \qq & \longmapsto & (\det u^{-1/n}\,u\, ,\,\det
u^{1/n})\\ 
ds\qq & \hbox{$\longleftarrow \!\!\! |$} & (\ s\ ,\ d\
)\ee define the isomorphism
\bb U(n)= SU(n)\times U(1)\ /\ \zz_n,\ee
where $\zz_n$ permutes the $n$ roots of the
determinant. For instance for $n=2$, we have
\bb \pp{i&0\cr 0&i}\ \longmapsto\ 
\left(\pp{+1&0\cr 0&+1},+i\right)\sim
\left(\pp{-1&0\cr 0&-1},-i\right).\ee
The map $\rho(u)=\det u^z\,u$, $u\in U(n),\ z\in\zz$,
defines a representation of $U(n)$. Under the above
isomorphism it induces the fundamental
representation of $SU(n)$ with $U(1)$ charges $1+zn$:
\bb \det u^z\,u=d^{zn}\, d^{1/n}s=d^{1+zn}\, s.\ee
The $U(n)$ representation $\rho(u)= \det u^z=d^{zn}$
induces the $SU(n)$ singlet representation with
$U(1)$ charges $zn$ and likewise for the adjoint
representation. The general theory \cite{or} tells us
these are the only $U(1)$ charges possible for the
fundamental, singlet and adjoint
$SU(n)$ representations induced from (continuous,
unitary) $U(n)$ representations. 

Let us recall the representations of the standard
model, we denote by $(m_2,6y,m_3)$ the tensor
product of a $m_2$ dimensional representation
under $SU(2)$  and a $m_3$ dimensional
representation under $SU(3)$ with hypercharge y.
Note that $U(1)$ charges only make sense after
multiplication with the coupling constant $g_1$ and
we have multiplied the conventional hypercharges
by 6 in order to make them integers. The fermion
representations are:
\bb 
&\pp{u\cr d}_L:\  (2,1,3)&\qq\pp{e\cr \nu}_L:\ 
(2,-3,1) \cr 
&\qq u_R:\  (1,4,3)&\cr 
&\qq\qq d_R:\  (1,-2,3) &\qq\qq\qq e_R:\ (1,-6,1),\eee 
the gauge bosons:
\bb
W^{\pm},\cos\theta_w\,Z+\sin\theta_w\,{\rm
photon}:& (3,0,1)&\cr 
-\sin\theta_w\,Z+\cos\theta_w\,{\rm
photon}:&(1,0,1)&\cr 
{\rm gluons}:&(1,0,8)&\eee
and the Higgs scalar:
\bb \varphi:\ (2,-3,1).\eee
Surprisingly, nature has chosen these nine
irreducible representations (nineteen
representations if we take into account all three
generations of quarks and leptons) such that they can
all be induced from $SU(2)\times U(3)$ and
simultaneously from
$U(2)\times SU(3)$. In other words, the internal group
of the standard model can be reduced to 
\bb S(U(2)\times U(3))=
\,\frac{SU(2)\times U(1) \times SU(3)}
{\zz_2\times \zz_3}\, .\ee

\section{Connes' point of view}

Connes \cite{connes} has generalized Riemannian
spaces to include an uncertainty principle. As in
quantum mechanics this uncertainty is coded in an
associative, noncommutative involution algebra $\aa$
and a representation $\rho$ of $\aa$ on a Hilbert
space $\hh$. In quantum mechanics $\aa$ is the
algebra of observables and $\hh=\lll^2(\rm
configuration\ space)$. In order to capture the metric
and in order to generalize differentiation and
integration to the noncommutative setting Connes
introduces a selfadjoint operator $\dd$ on the Hilbert
space, the `Dirac operator'. In even dimensional spaces one also
needs a `chirality', a unitary operator $\chi$ and in
real spaces one needs a `real structure', an
anti-unitary operator $J$. The five items, $\aa,\ \hh,\ 
\dd,\ \chi,\ J$ are called
spectral triple. They are supposed to satisfy axioms,
that are calibrated on Riemannian spin manifolds
$M$, with the {\it commutative} algebra of
differentiable functions on $M$, $\aa= {\cal
C}^\infty(M)$, $\hh$ is the space of square integrable
spinors on which a function acts by pointwise
multiplication, the Dirac operator is the genuine one,
$\dd=\ddd$, the chirality is $\chi=\gamma_5$ and the
real structure $J$ is charge conjugation. The axioms
are chosen such that there is a one-to-one
correspondence between commutative spectral triples
and Riemannian spin manifolds. 

Let us spell out the spectral triple for the zero
dimensional internal space of the standard model with
one generation of quarks and leptons:
\bb \aa=\hhh\op\cc\op M_3(\cc)\,\owns\,(a,b,c),\ee
\bb \hh=\hh_L\op\hh_R\op\hh_L^c\op\hh_R^c.
\ee
\bb \hh_L&=&
\left(\cc^2\ot\cc^3\right)\ \op\ 
\left(\cc^2\ot\cc\right), \\
\hh_R&=&\left((\cc\op\cc)\ot\cc^3\right)\ 
\op\ \left(\cc\ot\cc\right).\ee
The first factor denotes isospin, the second colour. We
choose the following basis of $\hh=\cc^{30}$: 
\bb
 \pp{u\cr d}_L,\ 
\pp{\nu_e\cr e}_L,\ 
\matrix{u_R,\cr d_R,}\   e_R,\ 
\pp{u\cr d}^c_L,\ 
\pp{\nu_e\cr e}_L^c,\ \matrix{u_R^c,\cr d_R^c,}\ 
e_R^c.\eee
The representation $\rho$ is defined by
\bb \rho(a,b,c):=\pp{\rho_{w}(a,b)&0\cr 0&
\bar\rho_{s}(b,c)} := 
\pp{\rho_{wL}(a)&0&0&0\cr 
0&\rho_{wR}(b)&0&0\cr 
0&0&{\bar\rho_{sL}(b,c)}&0\cr 
0&0&0&{\bar\rho_{sR}(b,c)}}\ee
with
\bb\rho_{wL}(a)&:=&\pp{
a\ot 1_3&0\cr
0&a&},\qq
\rho_{wR}(b)\ :=\ \pp{
B\ot 1_3&0\cr
0&\bar
b},\\ \cr &&
B:=\pp{b&0\cr 0&\bar b},
\\ \cr    
  \rho_{sL}(b,c)&:=&\pp{
1_2\ot c&0\cr
0&\bar b1_2},\qq
\rho_{sR}(b,c)\ :=\ \pp{
1_2\ot c&0\cr
0&\bar b}.   
\ee
The Dirac operator does not occur in the following
calculation, we indicate it for completeness,
\bb \dd=\pp{0&\mm&0&0\cr 
\mm^*&0&0&0\cr 
0&0&0&0\cr 
0&0&0&0},\qq\mm=\pp{
\pp{m_u&0\cr 0&m_d}\ot 1_3
&0\cr
0&\pp{0\cr m_e}}.\ee
We will need chirality and charge conjugation,
\bb \chi=\pp{-1_{8}&0&0&0\cr 0&1_{7}&0&0\cr
 0&0&-1_{8}&0\cr 0&0&0&1_{7} },\qq
 J=\pp{0&1_{15}\cr 1_{15}&0}\circ 
\ {\rm complex \ conjugation}.\ee
\medskip\noindent
Gauge invariance can be defined by
\begin{itemize}\item
the group of unitaries,
\bb U(\aa)=\left\{u\in\aa,\ uu^*=u^*u=1\right\}=
SU(2)\times U(1) \times
U(3),\ee
\item
the automorphism group $Aut(\aa)$. As locally
every automorphism $\varphi$ of a matrix algebra
$\aa$ is inner, $\varphi (a)=uau^{-1}$, for a unitary
$u\in U(\aa)$, we can obtain the automorphism
group as image of the unitary group under the
representation $\varphi$ of $U(\aa)$ on the vector
space $\aa$,
\bb Aut(\aa)\sim SU(2)\times SU(3),\ee
\item
the `covering' $Aut^\hh(\aa)$ of the automorphism
group on the Hilbert space $\hh$. The covering is
achieved by the {\it physical} representation of the
group of unitaries on the fermionic Hilbert space
$\hh$, $\rho(u)J\rho(u)J^{-1}$. This representation
is physical because it re-establishes invariance under
charge conjugation. Note that it is not an algebra
representation, it defines a bimodule. Its image is
locally
\bb Aut^\hh(\aa)\sim U(\aa)= SU(2)\times U(1)
\times  U(3).\ee 
\end{itemize}

After these unsuccessful attempts to obtain the
internal group of the standard model $G=SU(2)\times
U(1)\times SU(3)$, we are reduced to reduce the group
of unitaries by an {\it ad hoc} condition, the
`unimodularity' condition. At least it has the virtue to
make sense for general algebras $\aa$ \cite{har}.
For the finite dimensional algebra of the standard
model, the unimodularity condition can be formulated
conveniently on the Lie algebra level: 
\bb G\sim\exp\left\{X\in u(\aa),\  \t [P(\rho(X)+J\rho
(X) J^{-1})]=0\right\},\ee
where $P$ is the projection on the particles,
$\hh_L\op\hh_R$.
For the standard model, the unimodularity condition is
equivalent to the requirement that the physical
representation $\rho(u)J\rho(u)J^{-1}$ be free of
gauge and gravitational anomalies \cite{anom},
\bb &\t[P\chi(\rho(X)+J\rho(X) J^{-1})^3]=0&,\\
&\t[P\chi(\rho(X)+J\rho(X) J^{-1})]=0&.\ee

A natural question now is: can we modify the physical
representation of $U(\aa)$ by a phase, that is a central
element, such that the image of the modified
representation is locally isomorphic to $G$? Let us call
$\sigma$ the modified representation for which we
try the following ansatz:
\bb \sigma(u)=\rho(u)J\rho(u)J^{-1}\,
\rho(p(u))J\rho(p(u))J^{-1}, \qq u\in U(\aa),\ee
with the phase $p$,
\bb p:U(\aa)= SU(2)\times U(1) \times U(3)&
\longrightarrow& U(\aa)\cap {\rm
center}(\aa)\cr  u=(u_2,u_1,u_3)&\longmapsto& p(u)=
(1_2,u_1^\alpha\det u_3^\mu,u_1^\beta\det
u_3^\nu 1_3).\ee
$\sigma$ is a representation because $p(u\tilde
u)= p(u)p(\tilde u)$ and
\bb \sigma(u)=\rho(up(u))J\rho(up(u))J^{-1}.\ee
We will need the explicit form of $\rho\, J\rho
J^{-1}$ restricted to the particles,
\bb\rho(u_2,u_1,u_3)J\rho (u_2,u_1,u_3)J^{-1}=
\pp{u_2\ot u_3&0&0&0\cr 
0&u_2\,u_1^{-1}&0&0\cr 
0&0&\pp{u_1u_3&0\cr 0&u_1^{-1}u_3}&0\cr 
0&0&0&u_1^{-2}}.\ee
We want that the representation $\sigma$
restricted to the group $G$ of the
standard model close to identity
 coincide with the physical representation,
\bb \sigma(u)=\rho(u)J\rho(u)J^{-1},\qq
{\rm for \ all\ }u\in G,\qq u\sim 1,\ee
and that the image of $\sigma$ be locally isomorphic
to $G$,
\bb
\sigma(U(\aa))=\sigma(G)=\rho(G)J\rho(G)J^{-1}.\ee
Let us write down the infinitesimal form of $\sigma$.
It is sufficient to keep track of the two $U(1)$s. Since
$\sigma$ is invariant under charge conjugation we
restrict the computation to the particles:
\begin{equation}
\sigma\left(1_2,e^{i\theta_1},e^{i\theta_3}\,1_3\right)
=1_{15}+
i\ {\rm diag}\left(\begin{array}{c}
[\beta\theta_1+(1+3\nu)\theta_3]1_2\ot 1_3\\[2mm] 
-[(1+\alpha)\theta_1+3\mu\theta_3]1_2\\[2mm]
[(1+\alpha+\beta)\theta_1+(1+3\mu+3\nu)\theta_3
]1_3\\[2mm] 
[(-1-\alpha+\beta)\theta_1+(1-3\mu+3\nu)\theta_3
]1_3\\[2mm]
-2[(1+\alpha)\theta_1+3\mu\theta_3]\end{array}
\right)\
+O(\theta^2).\label{left} \end{equation}
For all values of $\theta_1$ et
$\theta_3$ this expression must be equal to the
exponential of hypercharge ,
\bb 1_{15}+i2\theta\,{\rm
diag}\pp{
y_1 1_2\ot 1_3\cr y_2 1_2\cr y_3 1_3\cr y_4 1_3\cr 
y_5}\ +O(\theta^2).\label{right}\ee
In our normalization the hypercharge values are, 
\bb y_1\ =&{\textstyle\frac{1}{6}},\qq &
6y_1=1\ {\rm mod}\ 2\qq{\rm and}\qq 1\ {\rm mod}\
3,\cr  
y_2\ =&-{\textstyle\frac{1}{2}},\qq &
6y_2=1\ {\rm mod}\ 2\qq{\rm and}\qq 0\ {\rm mod}\
3,\cr 
y_3\ =&{\textstyle\frac{2}{3}},\qq &
6y_3=0\ {\rm mod}\ 2\qq{\rm and}\qq 1\ {\rm mod}\
3,\cr 
y_4\ =&-{\textstyle\frac{1}{3}},\qq &
6y_4=0\ {\rm mod}\ 2\qq{\rm and}\qq 1\ {\rm mod}\
3,\cr 
y_5\ =&-1,\qq &
6y_5=0\ {\rm mod}\ 2\qq{\rm and}\qq 3\ {\rm mod}\
3.\label{cond}\ee
Equating the two expressions (\ref{left}) and
(\ref{right}) yields five equations. The last three 
equations are simply combinations of the first two
thanks to the three experimental identities,
$y_1-y_2=y_3$, $y_1+y_2=y_4$ and $2y_2=y_5$.
Note that if $y_1$ and $y_2$ satisfy the conditions
from $S(U(2)\times U(3))$ recalled in equations
(\ref{cond}), then this is also true for $y_3$, $y_4$
and $y_5$ computed with the three experimental
identities. We rewrite the first two equations as:
\bb
\pp{1+\alpha&3\mu\cr 
\beta&1+3\nu}\pp{\theta_3/\theta\cr 
\theta_1/\theta}=\pp{-2y_2\cr 2y_1}=
\pp{1\cr 1/3}.\label{det}\ee
In the physical representation $\rho J\rho J^{-1}$, 
$\alpha=\beta=\mu=\nu=0$, hypercharge is given
by the following linear combination of the two
$u(1)$s, 
$\theta_3={\textstyle\frac{1}{3}} \theta_1$, 
\bb u_{\rm hypercharge}=\left(1_2, e^{i\theta},
e^{i\theta/3}1_3\right).\ee
We want the two representations of hypercharge to
coincide,
\bb \sigma(u_{\rm hypercharge})=
\rho(u_{\rm hypercharge})J\rho(u_{\rm
hypercharge})J^{-1}.\ee
This is equivalent to $\alpha=-\mu$ and
$\beta=-\nu$. Furthermore we want the image of
the representation to be locally isomorphic to 
$G$. This is equivalent to a vanishing determinant of
the matrix in equation (\ref{det}),
\bb (1+\alpha)(1+3\nu)=\beta\,3\mu.\ee
This gives:
\bb \sigma(u_2,u_1,u_3)
=\rho(u_2,\det \tilde u_3,\tilde u_3)J\rho(u_2,\det
\tilde u_3,\tilde u_3)J^{-1},\qq \tilde u_3:= 
u_1^\beta \det u_3^{-\beta}\, u_3.
\ee
As a matter of fact, the modified representation
$\sigma$  ignores
$u_1$ and it is indeed a representation of $SU(2)\times
U(3)$.

\section{Conclusion}

The unimodularity condition remains a disturbing
feature in the geometric formulation of the standard
model. This condition is connected to
anomaly cancellation \cite{anom}, an intriguing
feature of the standard model. We remark that it is also
connected to another intriguing feature of the standard
model, namely that the internal group may be reduced
to $SU(2)\times U(3)$. Of course we would like to use
the reduction to $U(2)\times SU(3)$ as well. This
points towards the algebra $\aa= M_2(\cc)\op
M_1(\cc)\op M_3(\cc)$, that has been suggested by
Connes in the context of quantum groups \cite{tre}
and that is a major motivation of this workshop.
Within noncommutative Yang-Mills
theories, this algebra leads to an unacceptable light
neutral scalar
\cite{pris}. The phenomenological analysis of this
algebra within the spectral action is in progress
\cite{kra}.  

\bigskip

\noindent
{\it Acknowledgements:} We are greatly indepted to
Raymond Stora who raised the crucial issue of phases.

\vfil\eject

 \end{document}